\documentclass{elsart}

\usepackage{natbib}

 \usepackage{graphicx}

\usepackage{amssymb}
\usepackage{amsmath}

\begin{document}

\newcommand{\ETAL}{{\it et al.}}
\newcommand{\EG}{e.g.~}
\newcommand{\IE}{i.e.~}
\newcommand{\CF}{cf}

\newcommand{\CRIT}{{\rm crit}}
\newcommand{\TOT}{{\rm tot}}
\newcommand{\MAT}{{\rm m}}
\newcommand{\CDM}{{\rm c}}
\newcommand{\BAR}{{\rm b}}

\newcommand{\SCAL}{{\rm S}}

\newcommand{\EFF}{{\rm eff}}
\newcommand{\GAL}{{\rm g}}

\newcommand{\UUNIT}[2]{
{\;\mathrm{#1}^{#2}} }

\newcommand{\MIX}{{\rm MIX}}
\newcommand{\AD}{{\rm AD}}
\newcommand{\BI}{{\rm BI}}
\newcommand{\CI}{{\rm CI}}
\newcommand{\NID}{{\rm NID}}
\newcommand{\NIV}{{\rm NIV}}

\newcommand{\ep}{\epsilon}
\newcommand{\Ga}{\Gamma}
\newcommand{\OLa}{\Omega_{\Lambda}}
\newcommand{\La}{{\Lambda}}
\newcommand{\Om}{\Omega}
\newcommand{\om}{\omega}
\newcommand{\si}{\sigma}
\newcommand{\OO}{{\mathcal O}}
\newcommand{\FG}[1]{Fig.~#1}
\newcommand{\pars}{\mathbf{\Theta}}
\newcommand{\Li}{\mathcal{L}}
\newcommand{\PD}{\mathcal{P}_n}
\newcommand{\On}{\mathcal{SO}_n}
\newcommand{\pert}[1]{\frac{\delta \rho_{#1}}{\rho_{#1}}}

\newcommand{\be}{\begin{equation}}
\newcommand{\ee}{\end{equation}}
\def\gsim{\raise 0.4ex\hbox{$>$}\kern -0.7em\lower 0.62
ex\hbox{$\sim$}}
\def\lsim{\raise 0.4ex\hbox{$<$}\kern -0.8em\lower 0.62
ex\hbox{$\sim$}}
\newcommand{\lims}[2]{_{#1}^{#2}}

\begin{frontmatter}


 \title{The cosmological constant and the paradigm of adiabaticity}
 \author{Roberto Trotta}
 \ead{roberto.trotta@physics.unige.ch}
 \ead[url]{http://theory.physics.unige.ch/$\sim$trotta}
 \address{D\'epartement de Physique Th\'eorique, Universit\'e de
  Gen\`eve, 24 quai Ernest Ansermet, CH--1211 Gen\`eve 4, Switzerland}

\begin{abstract}
We discuss the value of the cosmological constant as recovered
from CMB and LSS data and the robustness of the results when
general isocurvature initial conditions are allowed for, as
opposed to purely adiabatic perturbations. The Bayesian and
frequentist statistical approaches are compared. It is shown that
pre-WMAP CMB and LSS data tend to be incompatible with a non-zero
cosmological constant, regardless of the type of initial
conditions and of the statistical approach. The non-adiabatic
contribution is constrained to be $\leq 40\%$ ($2\sigma$ c.l.).

\end{abstract}


\end{frontmatter}

\section{Introduction}

There are now at least 5 completely independent observations which
consistently point toward a majority of the energy-density of the
Universe being in the form of a ``cosmological constant'', $\OLa$.
Those observations are: cosmic microwave background anisotropies
(CMB), large scale structure (LSS), supernovae typ IA, strong and
weak gravitational lensing. The very nature of this mysterious
component remains unknown, and the so called ``smallness problem''
(i.e. why $\OO({\OLa}) \sim 1$ and not $\OLa \gsim 10^{58}$ as
expected from particle physics arguments) is still unsolved. It is
therefore important to test the robustness of results indicating a
non-vanishing cosmological constant with respect to non-standard
physics. One possibile extension of the ``concordance model'' is
given by non-adiabatic initial conditions for the cosmological
perturbations, \IE isocurvature modes. Another test is the use of
a different statistical approach then the usual Bayesian one,
namely the frequentist method. We discuss this points in the next
section, and present their application to the cosmological
constant problem in section 3. Section 4 is dedicated to our
conclusions.

\section{Testing the assumption of adiabaticity}
\subsection{Statistics}
Most of the recent literature on cosmological parameters
estimation uses {\it Bayesian inference}: the Maximum Likelihood
(ML) principle states that the best estimate for the unknown
parameters is the one which maximizes the likelihood function.
Therefore, in the grid-based method, one usually minimizes the
$\chi^2$ over the parameters which one is not interested in. Then
one defines $1 \si$, $2 \si$ and $3 \si$ likelihood contours
around the best fit point, as the locus of models within $\Delta
\equiv \chi^2 - \chi^2_{\rm ML} = 2.30$, $6.18$, $11.83$ away from
the ML value for the joint likelihood in two parameters, $\Delta =
1$, $4$, $9$ for the likelihood in only one parameter. Based on
Bayes' Theorem, likelihood intervals measure our degree of belief
that the particular set of observations used in the analysis is
generated by a parameter set belonging to the specified interval
\cite{statistics}. Since Bayesian likelihood contours are drawn
with respect to the ML point, if the best fit value for the
$\chi^2$ is much lower then what one would expect statistically
for Gaussian variables (\IE $\chi^2/F \approx 1$, were $F$ denotes
the number of degrees of freedom, dof), Bayesian contours will
underestimate the real errors.

The grid-based parameter estimation method can however be used for
a determination of true exclusion region ({\it frequentist
approach}). The Bayesian and frequentist methods can give quite
different errors on the parameters, since the meaning of the
confidence intervals is different. The frequentist approach
answers the question: What is the probability of obtaining the
experimental data at hand, if the Universe has some given
cosmological parameters? To the extent to which the $C_\ell$'s can
be approximated as Gaussian variables, the quantity $\chi^2$ is
distributed according to a chi-square probability distribution
with $F = N - M$ dof, where $N$ is the number of independent
(uncorrelated) experimental data points and $M$ is the number of
fitted parameters. Since the chi-square distribution, $P^{(F)}$,
is well known, one can readily estimate {\em confidence
intervals}, by finding the quantile of $P^{(F)}$ for the chosen (1
tail) confidence level. The so obtained exclusion regions do not
rely on the ML point. On the other hand, they are rigorously
correct only if the assumption of Gaussianity holds, and the
number of dof is precisely known. In general one should keep in
mind that frequentist contours are less stringent than likelihood
(Bayesian) contours.

\subsection{Dependence on initial conditions}

CMB anisotropies are sensitive not only to the matter-energy
content of the universe, but also to the type of initial
conditions (IC) for cosmological perturbations. Initial conditions
are set at very early times, and determining them gives precious
hints on the type of physical process which produced them. In the
context of the inflationary scenario, the type of IC is related to
the number of scalar fields in the very early universe and to
their masses. For instance, the simplest inflationary model,
namely with only one scalar field, predicts adiabatic (AD) initial
conditions. In this case, the initial density contrast for all
components (baryons, CDM, photons and neutrinos) is the same, up
to a constant:
$$
\pert{b} = \pert{c} = \frac{3}{4}\pert{\gamma} = \frac{3}{4}
\pert{\nu} \equiv \Delta_{AD} \qquad \text{(AD).}
$$
This excites a cosine oscillatory mode in the photon-baryon fluid,
which induces a first peak at $\ell \approx 220$ in the angular
power spectrum for a flat universe. Another possibility are CDM
isocurvature initial conditions. Then the total energy-density
perturbation vanishes (setting $\pert{b}=\pert{\nu}=0$ without
loss of generality):
$$
\pert{\TOT} = \pert{c} + \pert{\gamma} = 0  \qquad \text{(CDM
ISO)}
$$
and therefore the gravitational potential $\Psi$ is approximately
zero as well (``isocurvature''). CDM isocurvature IC excite a sine
oscillation, and the resulting first peak in the power spectrum is
displaced to $\ell \approx 330$. Generation of isocurvature
initial conditions requires the presence of (at least) a second
light scalar field during inflation. The observation of the first
peak at $\ell = 220.1 \pm 0.8$ \cite{Page03} has ruled out the
possibility of pure CDM isocurvature initial conditions. However,
a subdominant isocurvature contribution to the prevalent adiabatic
mode cannot be excluded.

Beside AD and CDM isocurvature, the complete set of IC for a fluid
consisting of photons, neutrinos, baryons and dark matter in
general relativity consists of three more modes~\cite{BMT}. These
are the baryon isocurvature mode (BI), the neutrino isocurvature
density (NID) and neutrino isocurvature velocity (NIV) modes.
Those five modes are the only regular ones, \IE they do not
diverge at early times. The NID mode can be understood as a
neutrino entropy mode, while the NIV consists of vanishing density
perturbations for all fluids but non-zero velocity perturbations
between fluids. The CDM and BI modes are identical, and therefore
it suffices to consider only one of them. In the most general
scenario, one would expect all four modes to be present with
arbitrary initial amplitude and arbitrary correlation or
anti-correlation, with the restriction that their superposition
must be a positive quantity. For simplicity we consider the case
where all modes have the same spectral index, $n_\SCAL$. The most
general initial conditions are then described by the spectral
index $n_\SCAL$ and a positive semi-definite $4 \times 4$ matrix,
which amounts to eleven parameters instead of two in the case of
pure AD initial conditions. More details can be found in
Refs.~\cite{TRD1, TRD2}. The CMB and matter power spectra for the
different types of initial conditions are plotted in
\FG~\ref{fig:power_spectra}.

\subsection{The matter power spectrum}

\begin{figure}[tb]
\begin{center}
\includegraphics[width=5.5cm]{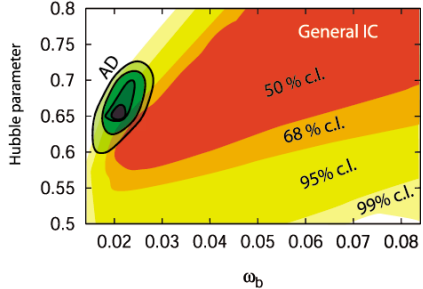}
\end{center}
 \caption{Joint Bayesian likelihood contours for the baryon density
 $\omega_b$ and the Hubble parameter $h$, using pre-WMAP CMB
 data only. The tighter contours (shades of green) assume purely
 AD initial conditions, the wider contours (yellow/shades of red)
 include general isocurvature IC (from Ref.~\cite{TRD1}).}
\label{fig:TRD1}
\end{figure}

\begin{figure}[tb]
\begin{center}
\includegraphics[width=3.0cm]{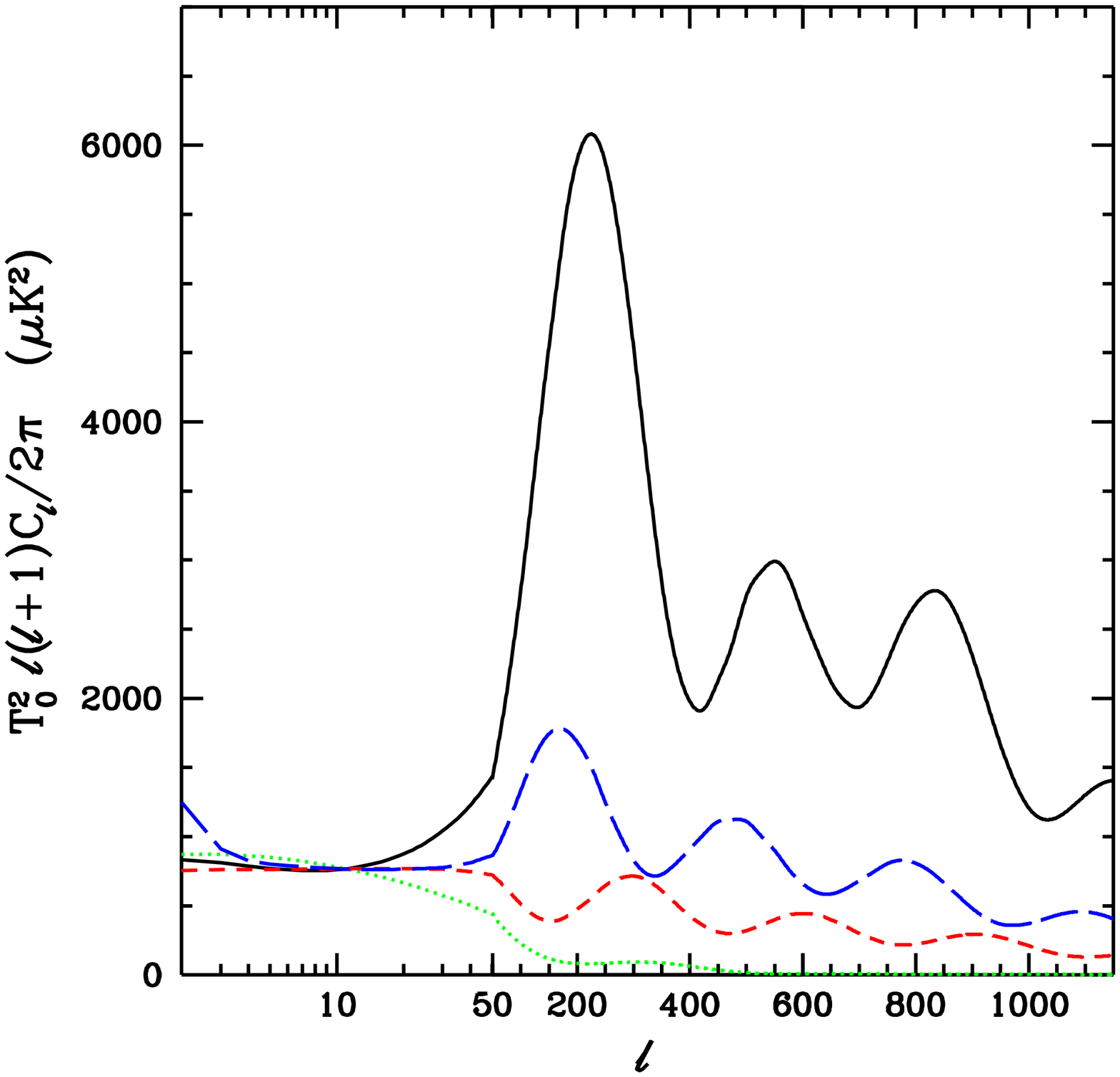}
\includegraphics[width=3.0cm]{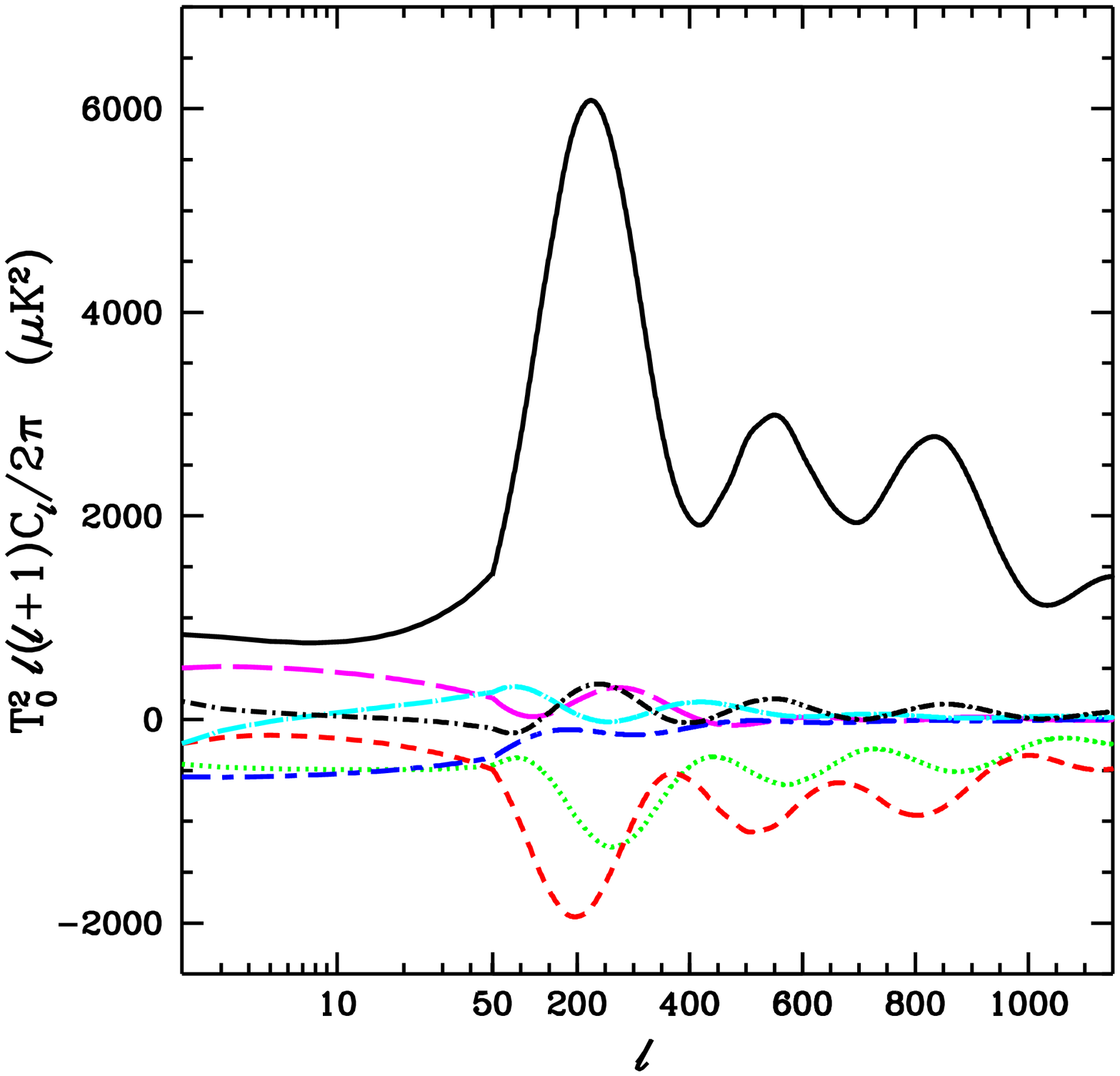}
\includegraphics[width=3.0cm]{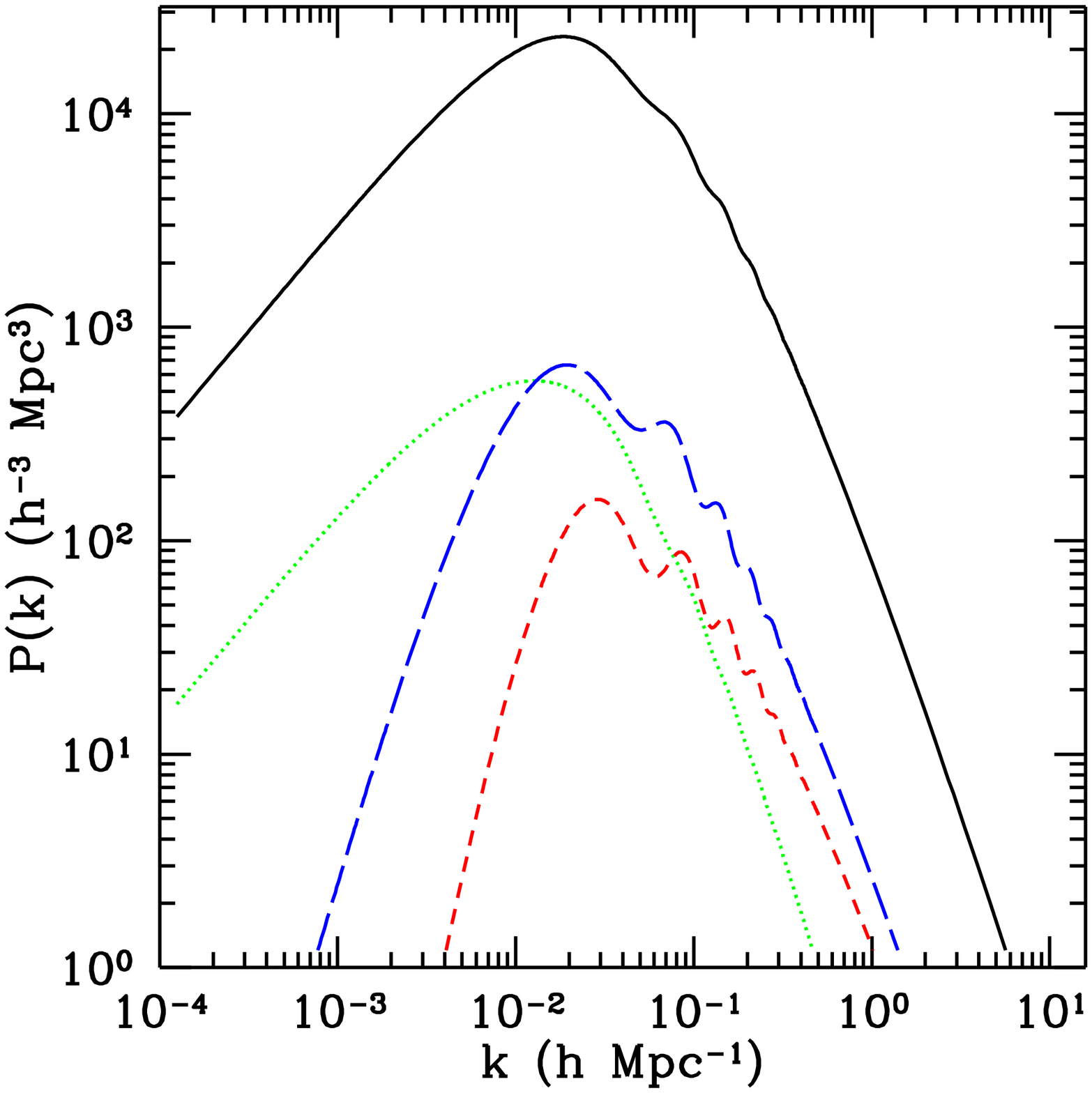}
\includegraphics[width=3.0cm]{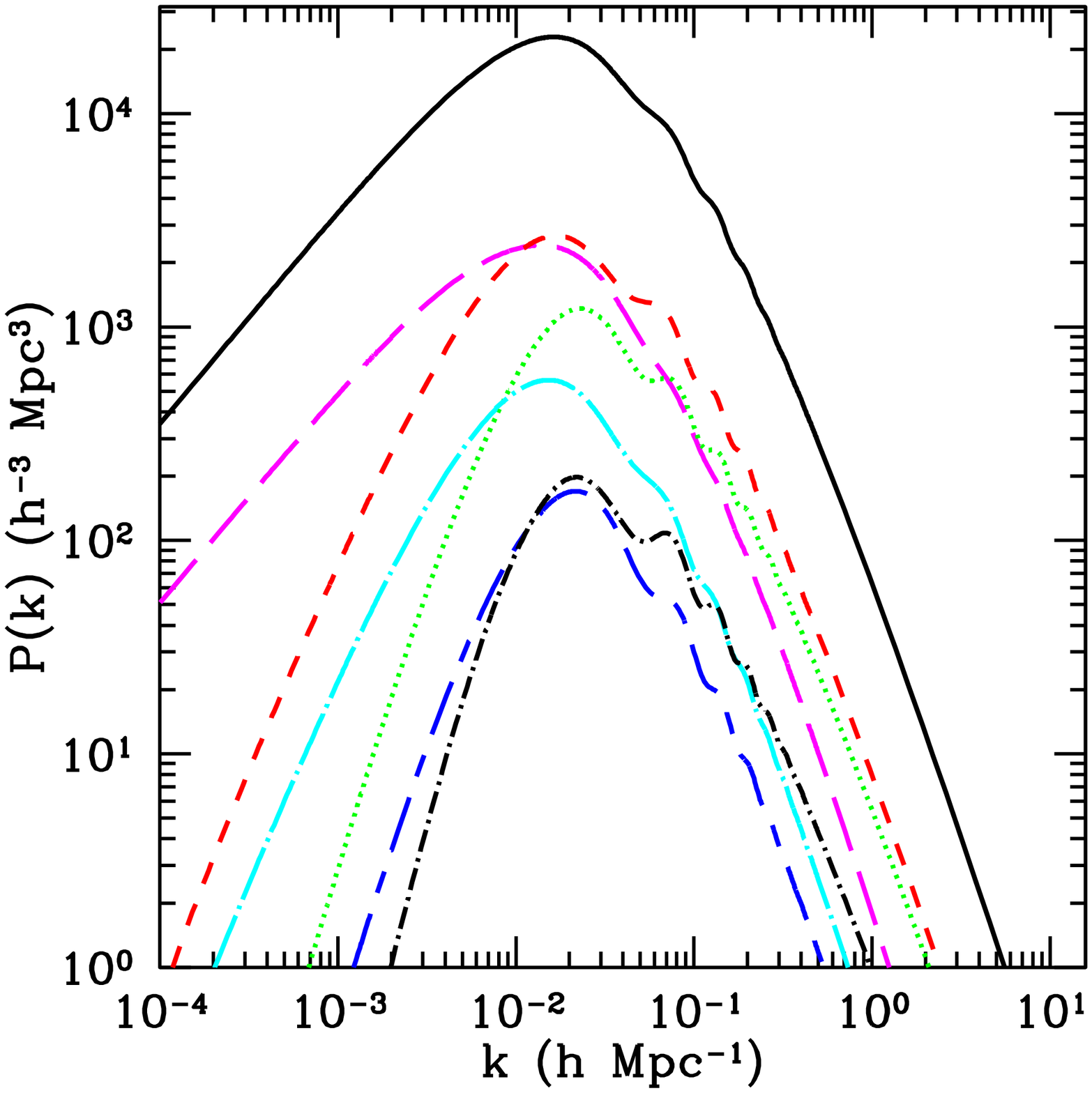}
\end{center}
 \caption{CMB (left) and matter (right) power spectra
 of the different auto- (odd panels) and
cross-correlators (even panels) for the standard $\Lambda$CDM
concordance model. The CMB power spectrum is COBE-normalized. The
color and line style codes are as follows:
 in the odd panels, AD: solid/black line, CI:
dotted/green line, NID: short-dashed/red line, NIV:
long-dashed/blue line;
 in the even panels, AD: solid/black line
(for comparison), $<\AD,\CI>$: long-dashed/magenta line,
$<\AD,\NID>$: dotted/green line, $<\AD,\NIV>$: short-dashed/red
line, $<\CI,\NID>$: dot-short dashed/blue line, $<\CI,\NIV>$:
dot-long dashed/light-blue line, and $<\NID,\NIV>$: dot-short
dashed/black line. }
\label{fig:power_spectra}
\end{figure}

Inclusion of general initial conditions in the analysis can lead
to very important degeneracies in the IC parameter space, which
spoil the accuracy with which other cosmological parameters can be
measured by CMB alone. This has been demonstrated in a striking
way for the case of the Hubble parameter and the baryon density in
Ref~\cite{TRD1}, \CF~\FG~\ref{fig:TRD1}. An effective way to break
this degeneracy is achieved by the inclusion of large scale
structure (LSS) data. The key point is that, once the
corresponding CMB power spectrum amplitude has been
COBE-normalized, the amplitude of the AD matter power spectrum is
nearly two orders of magnitude larger than {\it any} of the
isocurvature contribution (\CF~\FG~\ref{fig:power_spectra}).
Therefore the matter power spectrum essentially measures the
adiabatic part, and is nearly insensitive to isocurvature
contributions. The argument holds true for observations of the
matter spectrum on all scales, ranging from large scale structure
to weak lensing and Lyman $\alpha$-clouds. In view of optimally
constraining the isocurvature content, it is therefore essential
to combine those observations with CMB data, in order to break the
strong degeneracy among initial conditions which is present in the
CMB power spectrum alone \cite{Tprep}.

\section{The cosmological constant and isocurvature IC}

We apply the above statistical (Bayesian or frequentist) and
physical (general initial conditions, matter power spectrum)
considerations to the study of the cosmological constant problem
from pre-WMAP data. We outline the method and the main results
below (see Ref.~\cite{TRD2} for more details) and comment at the
end on the qualitative impact of the new WMAP data on those
findings.

Our analysis makes use of the COBE, BOOMERanG and Archeops data
\cite{CMBdata}, covering the range $3 \leq \ell \leq 1000$ in the
CMB power spectrum. For the matter power spectrum, we use the
galaxy-galaxy linear power spectrum from the 2dF data
\cite{2dFdata}, and we assume that light traces mass up to a
(scale independent) bias factor $b$, over which we maximise. The
main focus being on the type of initial conditions, we restrict
our analysis to only 3 cosmological parameters: the scalar
spectral index, $n_S$, the cosmological constant $\OLa$ in units
of the critical density and the Hubble parameter, $H_0 = 100 \,h
\UUNIT{km}{} \UUNIT{s}{-1} \UUNIT{Mpc}{-1}$.  We consider flat
universes only and neglect gravitational waves.

When we set to zero the isocurvature modes, we recover the
well-known results for purely AD perturbations. Because of the
``geometrical degeneracy'', CMB alone cannot put very tight lower
limits on $\OLa$ even if we allow only for flat universes. The
degeneracy can be broken either by putting an external prior on
$h$ or via the LSS spectrum, since $P_\MAT$ is mainly sensitive to
the shape parameter $\Gamma \equiv \Om_\MAT h$. Combination of CMB
and LSS data yields the following likelihood (Bayesian) intervals
for $\OLa$:
\begin{equation}
\OLa = 0.70 \lims{-0.05}{+0.05} \mbox{ at $1 \si$ $\quad$ and
$\quad$}
            \lims{-0.27}{+0.15}
\mbox{ at $3 \si$}.
\end{equation}

From the Bayesian analysis, one concludes that CMB and LSS
together with purely AD initial conditions require a non-zero
cosmological constant at very high significance, more than $7 \si$
for the points in our grid! However, our best fit has a reduced
chi-square $\chi/F= 0.59$, significantly less than $1$. This leads
to artificially tight likelihood regions: the observationally
excluded part of parameter space is less extended, and is given by
the frequentist analysis. From the frequentist approach, we obtain
instead the following confidence intervals:
\begin{equation}
0.15 < \OLa < 0.90 \mbox{ at $1 \si$ $\quad$
 and $\quad$}
 \OLa < 0.92 \mbox{ at $3 \si$}.
\end{equation}

\begin{figure}[tb]
\begin{center}
\includegraphics[width=5.5cm]{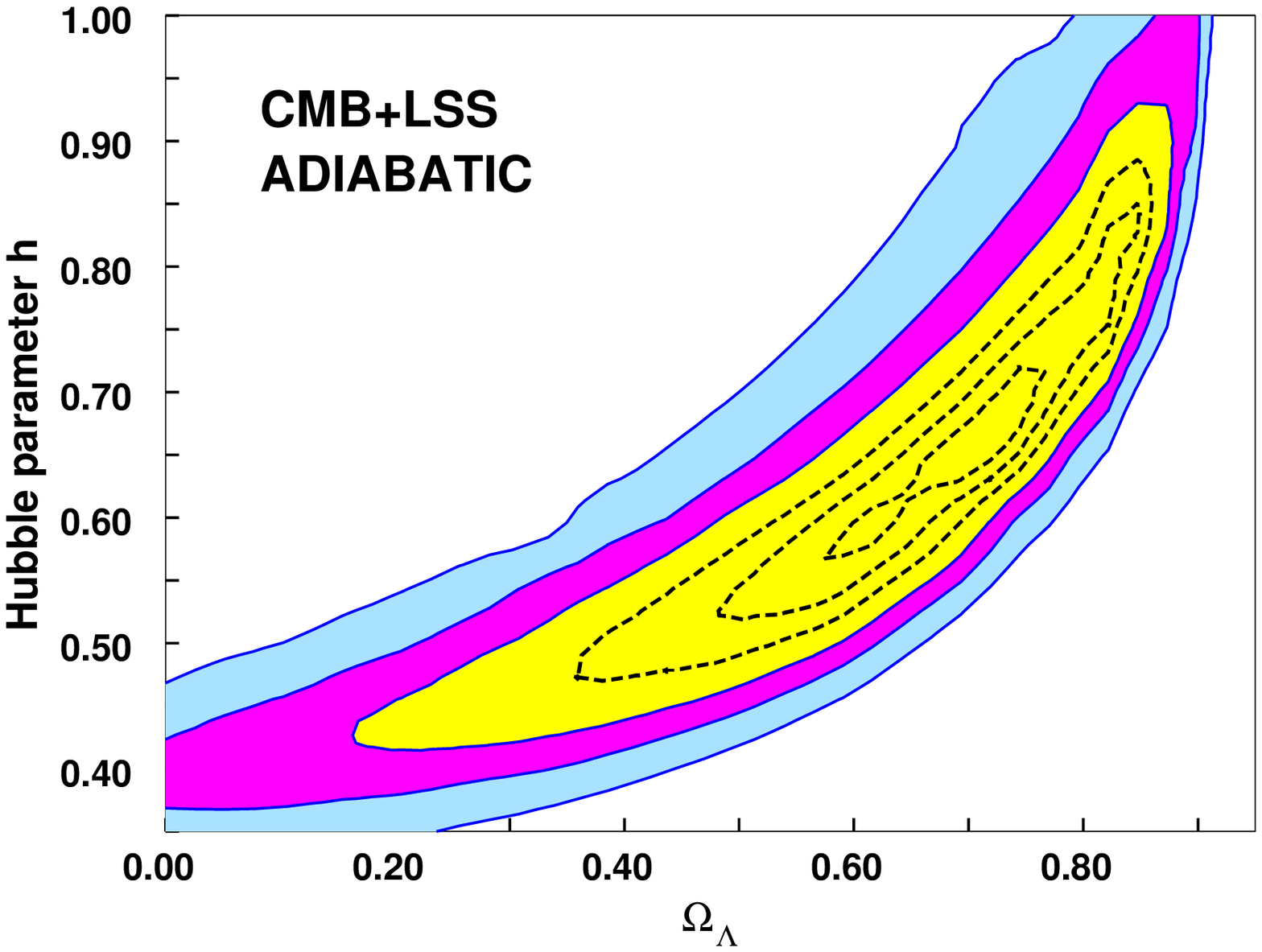}
\includegraphics[width=5.5cm]{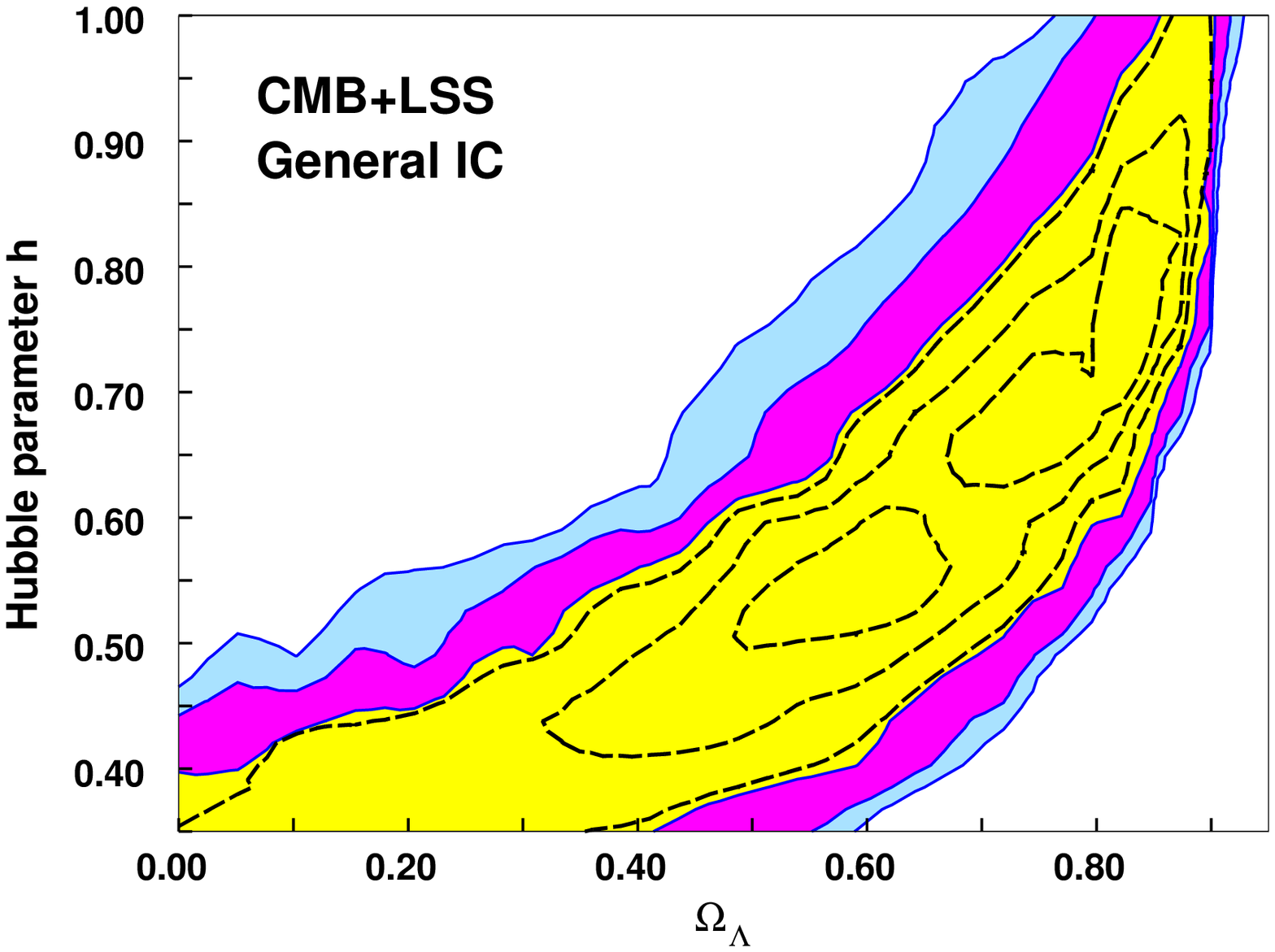}
\end{center}
 \caption{Bayesian (dashed lines) and
 frequentist (solid, filled) joint $1 \si$, $2 \si$, $3 \si$ contours using pre-WMAP CMB and 2dF data.
 The left panel assumes purely
 adiabatic IC, the right panel includes general isocurvature IC.}
\label{fig:ADGI}
\end{figure}

When we enlarge the space of models by including all possible
isocurvature modes, likelihood (Bayesian) and confidence
(frequentist) contours widen up along the $\OLa$, $h$ degeneracy,
and this produces a considerable worsening of the likelihood
limits. For general initial conditions we now find (Bayesian, CMB
and LSS together):
\begin{equation}
\OLa = 0.70 \lims{-0.10}{+0.15} \mbox{ at $1 \si$ $\quad$ and
$\quad$}
            \lims{-0.48}{+0.25}
\mbox{ at $3 \si$}.
\end{equation}
Again, the frequentist statistics give less tight bounds:
\begin{equation}
\OLa < 0.90 \mbox{ at $1 \si$ $\quad$
 and $\quad$}
 \OLa < 0.95 \mbox{ at $3 \si$},
\end{equation}
and in particular we cannot place any lower limit on the value of
the cosmological constant. A complete discussion can be found in
Ref.~\cite{TRD2}. Joint likelihood contours for $\OLa$, $h$ with
AD and general isocurvature initial conditions are plotted in \FG
\ref{fig:ADGI} for both statistical approaches.

\begin{figure}[tb]
\begin{center}
\includegraphics[width=5.5cm]{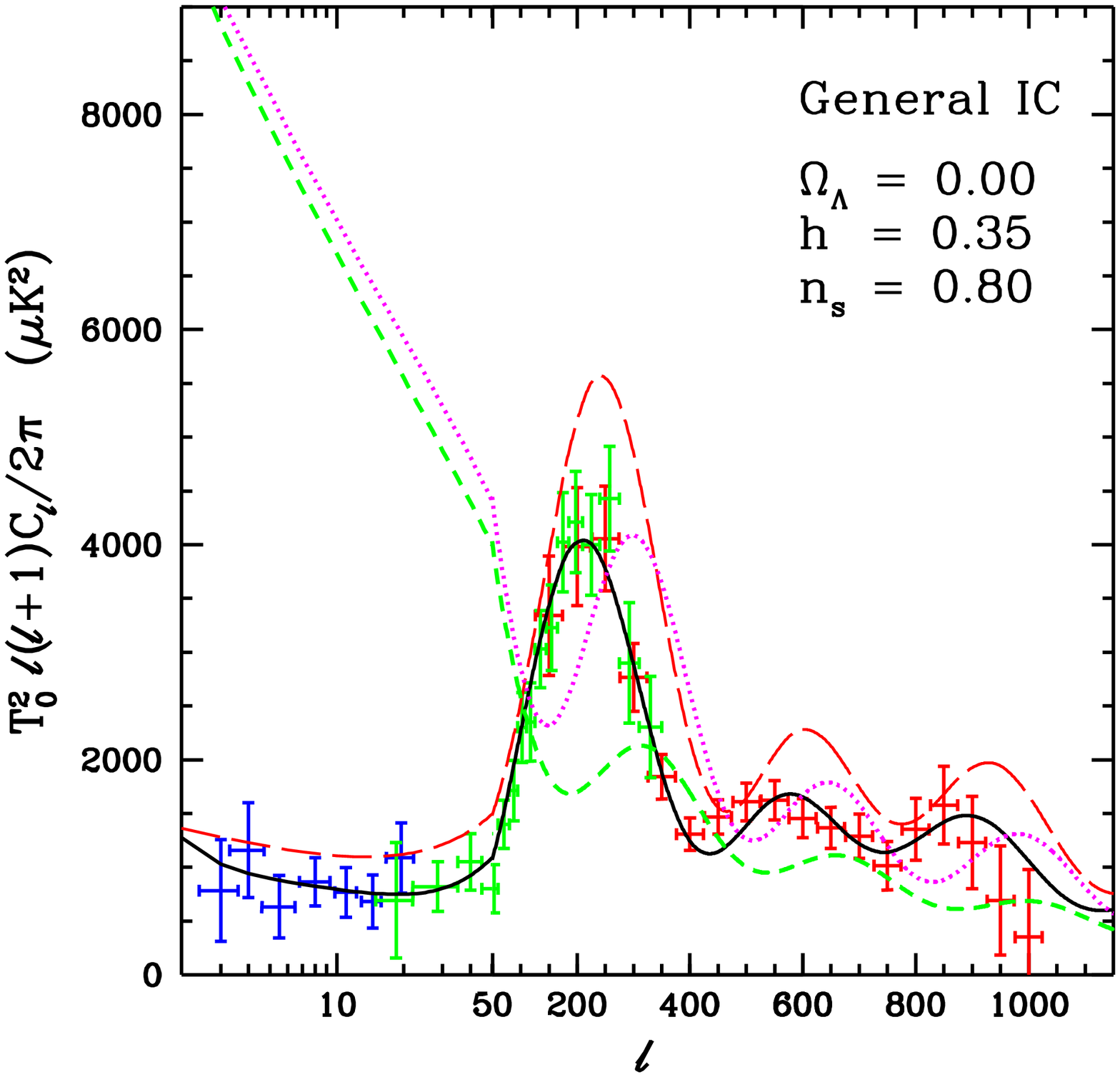}
\includegraphics[width=5.5cm]{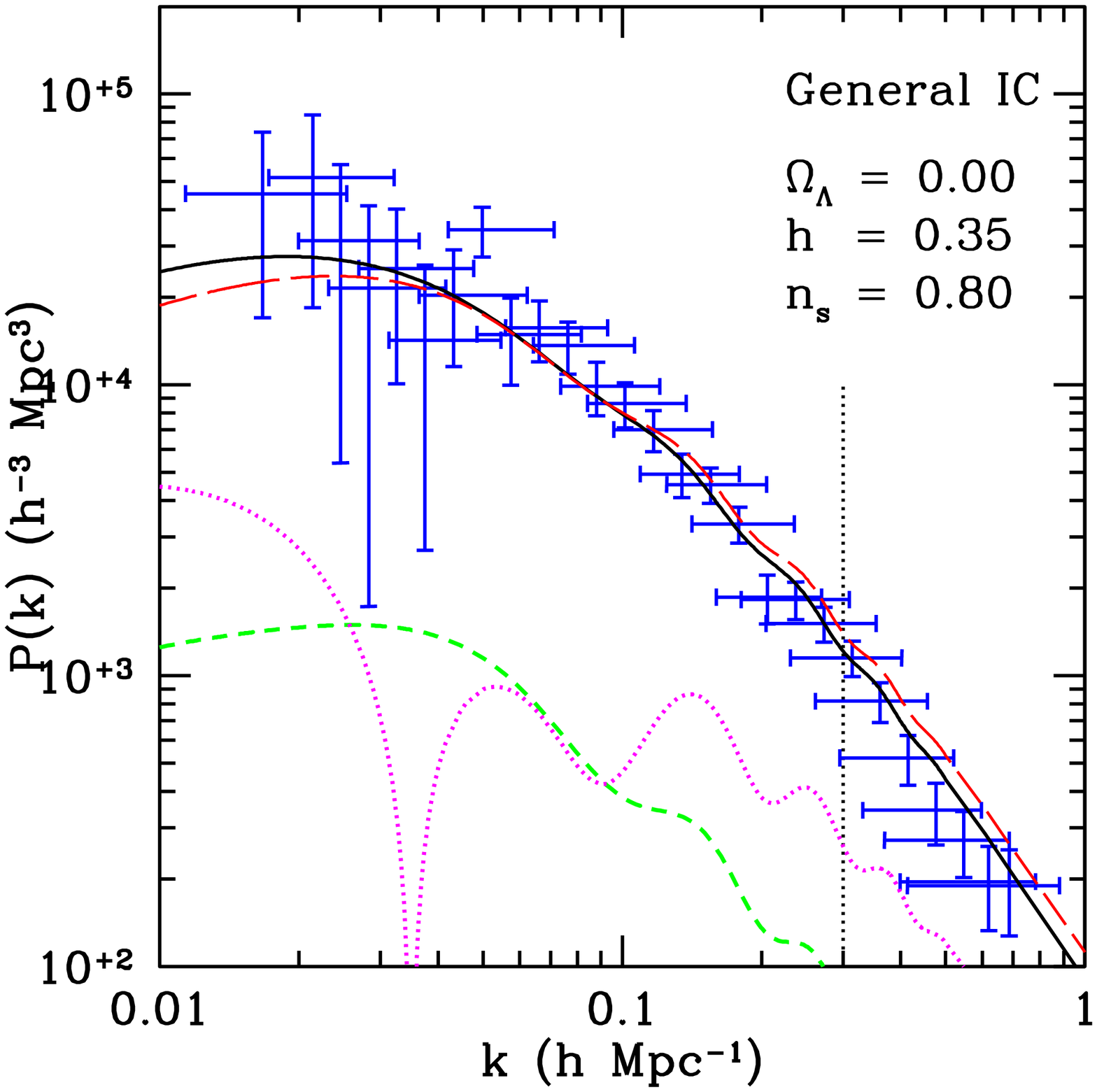}
\end{center}
 \caption{Best
fit with general IC and $\OLa = 0$, combining pre-WMAP CMB (left)
and 2dF (right) data. In both panels solid/black is the total
spectrum, long-dashed/red the purely AD contribution,
short-dashed/green the sum of the pure isocurvature modes,
dotted/magenta the sum of the correlators (multiplied by $-1$ in
the left panel and in absolute value in the right panel).}
\label{fig:AM_OL0}
\end{figure}

From the frequentist point of view, the region in the $\OLa, h$
plane which is incompatible with data at more than $3 \si$ is
nearly independent on the choice of initial conditions (compare
left and right panel of \FG \ref{fig:ADGI}). Enlarging the space
of initial conditions seemingly does not have a relevant benefit
on fitting pre-WMAP data with or without a cosmological constant.
In \FG \ref{fig:AM_OL0} we plot the best fit  model (which has
$\chi/F= 0.67$) with general initial conditions and $\OLa = 0$. As
a consequence of the red spectral index ($n_S=0.80$) and of the
absence of the early Integrated Sachs-Wolfe effect (since $\OLa
=0$), the best fit model has a very low first acoustic peak, even
in the presence of isocurvature modes. This is compatible with the
BOOMERanG and Archeops data only if the absolute calibration of
the experiments is reduced by $28\%$ and $12\%$, respectively.
Furthermore, this best fit model has a rather low value of the
Hubble parameter, $h=0.35$, which is many sigmas away from the
value obtained by the HST Key Project, namely $h=0.72 \pm 0.08$
\cite{HST}. We conclude that a good fit to the pre-WMAP CMB data
combined with LSS measurements can only be obtained at the price
of pushing hard the other parameters, even when general initial
conditions are allowed for.

\begin{figure}[tb]
\begin{center}
\includegraphics[width=5.5cm]{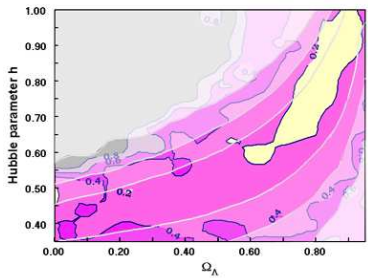}
\end{center}
\caption{Isocurvature content $0.0 \leq \beta \leq 1.0$ of best
fit models with pre-WMAP CMB and 2dF data. The contours are for
$\beta = 0.20, 0.40, 0.60, 0.80$ from the center to the outside.
Shaded regions represents 1 to 3 $\sigma$ c.l..}
\label{fig:BETA_SHADE}
\end{figure}

Finally, in order to constrain deviations from perfect
adiabaticity, it is interesting to limit quantitatively the
isocurvature contribution. To this end, one can phenomenologically
quantify the isocurvature contribution to the CMB power by a
parameter $0 \leq \beta \leq 1$, defined in Ref.~\cite{TRD2}, so
that purely AD IC are characterized by $\beta=0$, while purely
isocurvature IC correspond to $\beta=1$. In
Fig.~\ref{fig:BETA_SHADE} we plot the value of $\beta$ for the
best fit models, with the frequentist exclusion regions
superimposed. Within $2\sigma$ c.l. (frequentist), the
isocurvature contribution to the IC is bounded to be less then
$40\%$.

Although a quantitative analysis using the more precise WMAP data
has not yet been carried out, some qualitative features of the
expected results can be discussed. In particular, the first peak
has been measured by WMAP to be 10\% higher then in previous
observations \cite{WMAP}. On the other hand, our work indicates
that the first peak is very suppressed even in the presence of
general IC for $\OLa=0$. Therefore one expects that WMAP data will
exclude with much higher confidence a vanishing cosmological
constant. In fact, our pre-WMAP best fit $\OLa=0$ model, when
compared to the WMAP data \cite{WMAP}, has $\chi^2_{WMAP}/F
\approx 4.4$, and is therefore found to be totally incompatible
with the new data. Furthermore, the constraints on non-adiabatic
contributions should improve considerably, especially in view of
the inclusion of polarization data \cite{BMTpol}.

\section{Conclusions}

We have shown that the statistical approach (Bayesian or
frequentist) can have an important impact in the determination of
errors from CMB and LSS data. We found that structure formation
data tend to prefer a non-zero cosmological constant even if
general isocurvature initial conditions are allowed for. The
isocurvature contribution is constrained to be $\leq 40\%$ at
$2\sigma$ c.l. (frequentist).

\section*{Aknowledgments}

It is a pleasure to thank Alessandro Melchiorri and all the
organizers of the workshop. I am also grateful to Alain Riazuelo
and Ruth Durrer for a most pleasant collaboration. RT is partially
supported by the Schmidheiny Foundation, the Swiss National
Science Foundation and the European Network CMBNET.

 \end{document}